\documentclass[10pt,twocolumn]{article}
\pdfoutput=1

\usepackage{graphics, graphicx}
\usepackage{amsmath}
\usepackage{epsfig}
\usepackage{latexsym}

\begin{document}

\twocolumn[
  \begin{@twocolumnfalse}
  \hrule \vspace{.4cm}
\begin{centering} {\Huge\bf Two-dimensional tile displacement\\} \end{centering}

\begin{centering} {\Huge\bf can simulate cellular automata\\} \end{centering}

\vspace{.5cm}
\begin{centering} {\large\bf Erik Winfree$^{1,2,3 \star}$ and Lulu Qian$^{1,2,3 \star}$\\} \end{centering}

\vspace{.3cm}
\begin{centering} {\it $^1$Bioengineering, $^2$Computer Science, $^3$Computation and Neural Systems\\
California Institute of Technology, Pasadena, CA 91125, USA \\
$^\star$e-mail: winfree@caltech.edu, luluqian@caltech.edu \\} \end{centering}

\begin{quotation} \noindent
Tile displacement is a newly-recognized mechanism in DNA nanotechnology that exploits principles analogous to toehold-mediated strand displacement but within the context of self-assembled DNA origami tile arrays.
Here, we formulate an abstract model of tile displacement for the simplest case: individual assemblies interacting with monomer tiles in solution.
We give several constructions for programmable computation by tile displacement, from circuits to cellular automata, that vary in how they use energy (or not) to drive the system forward (or not), how much space and how many tile types they require, and whether their computational power is limited to PTIME or PSPACE with respect to the size of the system.
In particular, we show that tile displacement systems are Turing universal and can simulate arbitrary two-dimensional synchronous block cellular automata, where each transition rule for updating the state of a $2 \times 2$ neighborhood is implemented by just a single tile.
\end{quotation}

\begin{centering} {\bf Keywords: DNA origami, tile displacement, cellular automata, reversible computation\\} \end{centering}

   \vspace{.4cm} \hrule \vspace{.4cm}

  \end{@twocolumnfalse}
  ]

\newcommand{\mysection}[1] {\vspace{.2cm}  \noindent {\bf \uppercase{#1}} \vspace{.2cm} }
\mysection{1. Introduction}

\noindent
A guiding principle in theoretical computer science has been ``mechanism-to-model'' exploration of connections between physical implementation and computational capabilities.
For example, what can be computed by systems of AND gates and OR gates is strictly less than what can be computed by systems of NOR gates, which in turn is less than what can be computed by finite state machines coupled with an unbounded memory tape~\cite{savage1998models}.
Likewise, molecular programming theory aims to understand how fundamental molecular mechanism can be used to build systems, and how the choice of mechanism determines the range of what can be built.
An example would be the self-assembly of molecular structures by programmable cooperative binding, which can reliably grow structures that cannot grow reliably via non-cooperative binding~\cite{doty2012theory}.
When a new molecular mechanism is discovered, it is of interest to understand the nature -- the limitations and capabilities -- of systems that exploit that mechanism.
Doing so entails formulation of an abstract model that captures the essential features of the mechanism, which can then be rigorously analyzed.

\begin{figure*}[tb!]
\centerline{\includegraphics[scale=0.3]{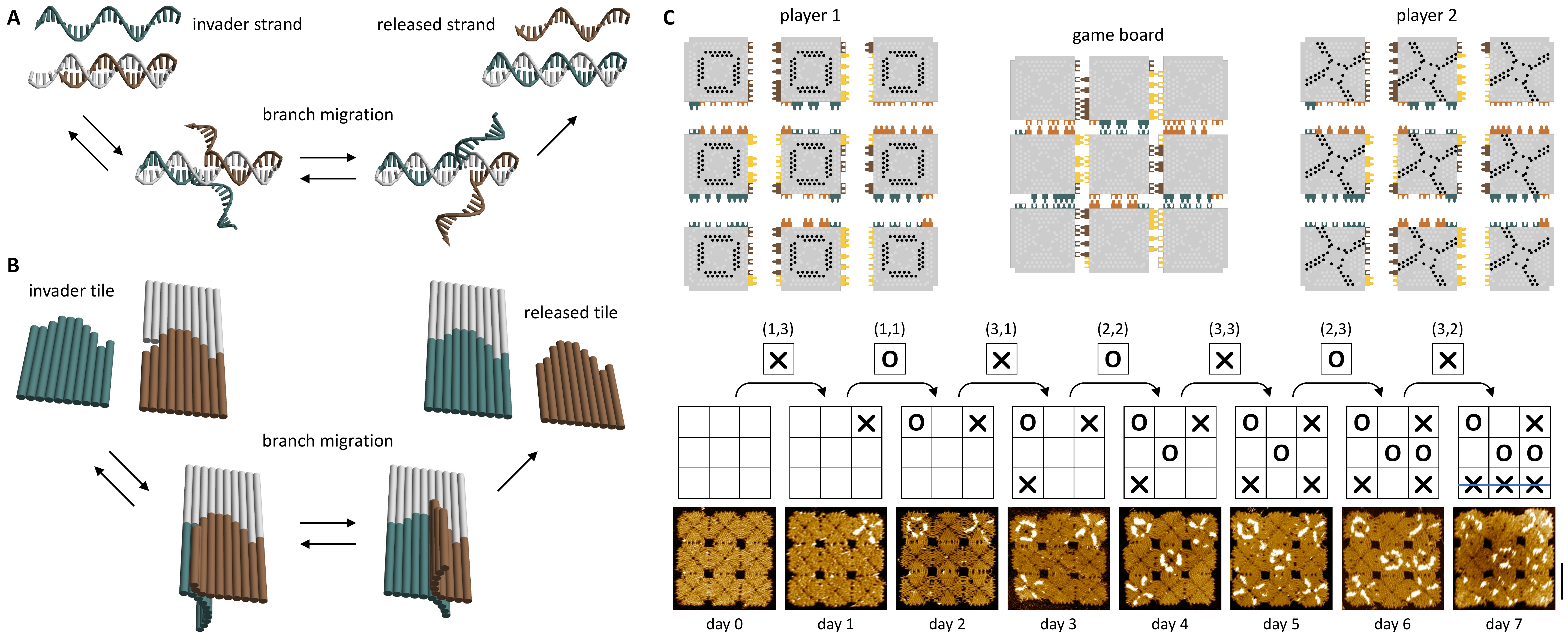}}
  \caption{\small {\bf }
  {\bf (A)} Strand displacement mechanism.  For scale, the DNA molecules are roughly 2~nm in diameter and 7~nm long.
  {\bf (B)} Tile displacement mechanism.  These hypothetical 10-helix DNA origami tiles are smaller than the 22-helix square tiles from~\cite{petersen2018information}.
  {\bf (C)} A tic-tac-toe game implemented using tile displacement (adapted from reference~\cite{petersen2018information}). Scale bar for atomic force microscopy (AFM) images is 100~nm.
  }
  \label{fgr:intro}
\end{figure*}

Since its invention two decades ago~\cite{yurke2000dna}, toehold-mediated DNA strand displacement has been a central mechanism for programming dynamical function in DNA nanotechnology~\cite{seeman2015structural,zhang2011dynamic}.
As shown in figure~\ref{fgr:intro}A, a stable complex of two strands can be reconfigured such that an invading strand replaces the original partner via a branch migration process.
The single-stranded portion of the original complex -- known as the toehold because that is where the invading strand initiates contact -- is critical for the displacement: an invader that does not match and bind to the toehold may be a million-fold slower to perform the displacement, and thus neglected as ``leak''.
In abstract models that consider networks of more complex (but still non-pseudoknotted) DNA molecules that interact in solution using toehold-mediated strand displacement reactions -- including both the irreversible mechanism shown here and a reversible variant known as ``toehold exchange'' -- have been shown capable of simulating arbitrary formal chemical reaction network dynamics and even Turing-universal computation~\cite{phillips2009programming,badelt2020domain,soloveichik2010dna,cardelli2013two,qian2010efficient,lakin2011modelling,yahiro2016implementation}.
However, a limitation of these results is that they are intrinsically distributed computations, where state is encoded within a collection of molecules in solution, and therefore a single test tube can perform only one computation at a time.
More complex molecular mechanisms, such as the hypothetical polymer-modifying enzymes envisioned by Bennett~\cite{bennett1982thermodynamics}, are in principle capable of performing independent Turing-universal computation in parallel in the same test tube.

The mechanism of ``tile displacement'', shown in figure~\ref{fgr:intro}B, was recently discovered during investigations into why the self-assembly of DNA origami tiles~\cite{rothemund2006folding,liu2011crystalline,tikhomirov2017fractal} failed to become kinetically trapped in undesired intermediates that the naive theory predicted~\cite{petersen2018information}.
There is a strong analogy to toehold-mediated strand displacement.
Beyond using components that are two orders of magnitude larger than the individual strands involved in strand displacement, the tile displacement mechanism has several distinct features.
(1) Nucleotides are on multiple helices that are oriented orthogonally to the axis of branch migration, rather than being on a single helix that is parallel to (identical to) the axis of branch migration.
(2) Tile-tile binding may be due to multiple helix-end stacking bonds~\cite{woo2011programmable} in addition to (or instead of) being due to base-pair formation.
%
%
(3) Toehold and branch migration domain specificity can be encoded both by tile geometry and by the sequences in multiple very short (1 or 2~nt) sticky ends, rather than being exclusively encoded by nucleotide sequences within a single helix.
(4) The released tile will be less flexible than a single-stranded oligonucleotide.
(5) Rather than having just one ``side'' and initiating displacement via a single toehold, tiles may have many (e.g. four) sides and may initiate displacement via cooperative action of multiple toeholds, as highlighted by the replacement of the central tile of a $3 \times 3$ tic-tac-toe game board~\cite{petersen2018information} shown in figure~\ref{fgr:intro}C.
Despite these differences, it remains that tile displacement is highly sensitive to toehold and branch migration sequences, such that the kinetics of tile displacement without a matching toehold may be orders of magnitude slower and similarly negligible as ``leak''.
Indeed, systems of interacting tile monomers and tile assemblies were shown to be reconfigurable by toehold-mediated tile displacement~\cite{petersen2018information}, and the same or similar constructs ought to be sufficient to implement more complex information-processing networks following, for example, the seesaw motif for circuits~\cite{qian2011simple} or the two-domain scheme for formal chemical reaction network dynamics~\cite{cardelli2013two}.

Here we are interested in whether the tile displacement mechanism enabled new ways of programming dynamical behaviors, beyond simply replicating strand displacement on a larger scale.
The ability to perform displacement within a two-dimensional array being an especially novel feature of tile displacement, we ask whether -- unlike existing strand displacement constructions -- reconfiguration of a single tile assembly in a constant soup of monomer tiles might be sufficient for substantial computation, in which case parallel computation could be achieved with each tile assembly performing an independent computation.
We present three results.
First, with a feedforward Boolean circuit laid out on the initial array, there is a tile set that, via displacement, propagates signals along wires and executes the specified logic.  This system is powered by the energy of toehold formation; the final state is in an energy minimum and cannot be reused.
Second, as a simplification and generalization of the first construction, any one-dimensional cellular automaton can be directly translated into a set of tiles such that a wave of tile displacement converts an assembly, initially empty but for the input, into the space-time history of the cellular automaton. This system is powered by a concentration difference between the monomer tiles that are invading over those that are displaced.
The above two constructions displace each tile in the original array at most once, using energy that is linear in the area used.
The third construction addresses whether iterated computation can be performed in-place, which requires replacing the tile at a given location an unbounded number of times.
Remarkably, using locally reversible asynchronous tile displacement, we can simulate arbitrary synchronous block cellular automata that use the $2 \times 2$ Margolus neighborhood, including his globally reversible Billiard Ball Model that is known to be Turing universal by simulation of infinite or finite recurrent Boolean circuits~\cite{margolus1984physics,toffoli1987cellular,margolus1999crystalline}.
A key issue is how to bias the computation forward; we show that it is enough to include a large empty part of the array into which entropy is injected.

This paper does not aim for novel advances in molecular programming that make technical applications closer to reality.  We are sharing these observations mainly because we find them to be beautiful and surprising.  Tile displacement may indeed be useful for reconfiguration of adaptive molecular systems, but for most implementation goals that are merely computational, there are more direct and more reliable ways to achieve them using other mechanisms in DNA nanotechnology.  However, it is remarkable that a molecular mechanism accidentally discovered in the laboratory gives rise to a theoretical model with such natural and direct connections to an esoteric but well-studied model of computation that arose in the study of the fundamental physics and ballistic motion.  We hope you will see through our imperfect figures and clumsy explanations to see the poetry within the concepts~\cite{winfree1964poets}.

\begin{figure*}[tb!]
\centerline{\includegraphics[scale=0.3]{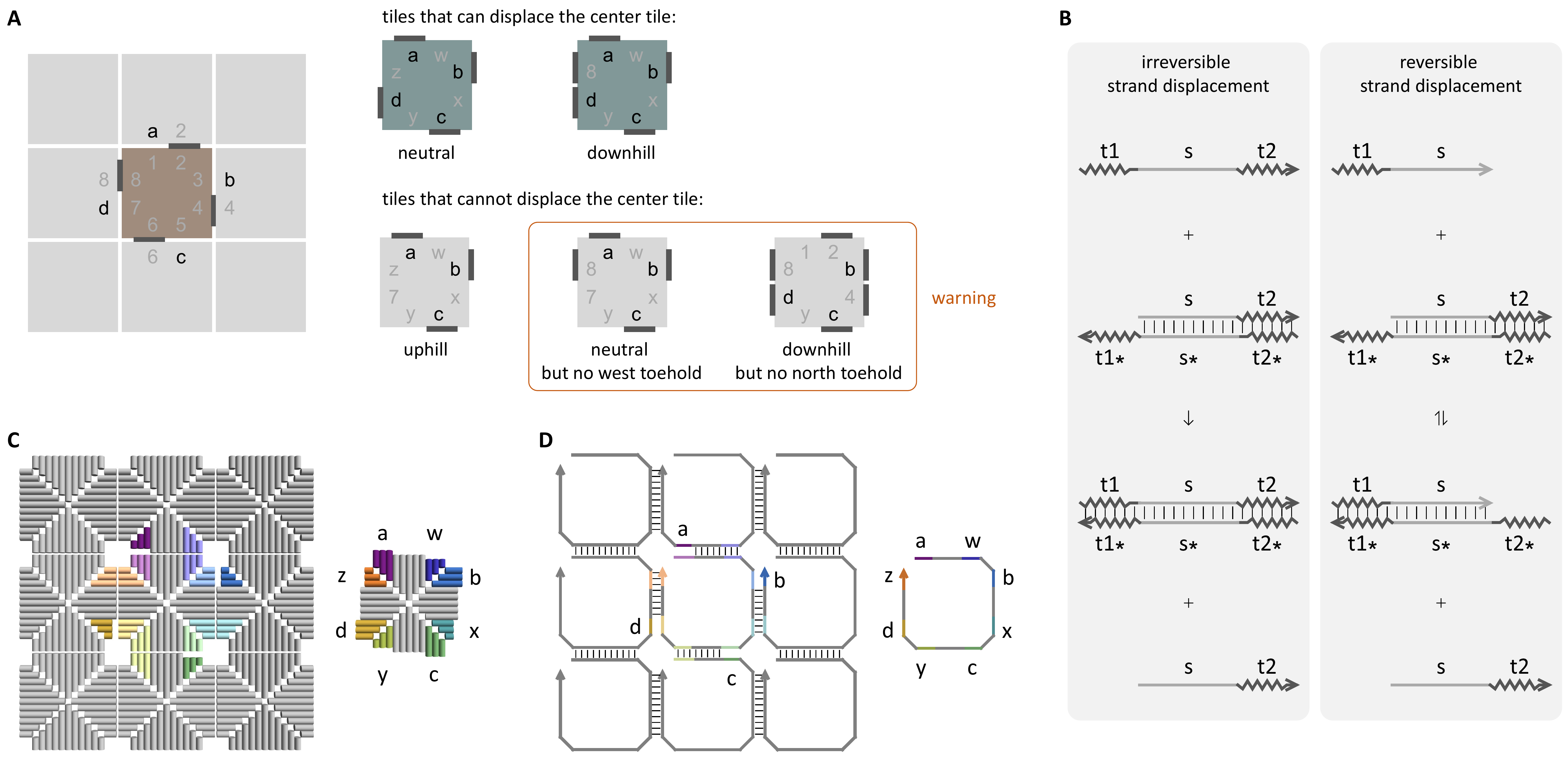}}
  \caption{\small {\bf }
  {\bf (A)} Abstract tile displacement model.
  {\bf (B)} Irreversible and reversible strand displacement.
  {\bf (C)} DNA origami tile implementation.
  {\bf (D)} Single-stranded tile implementation.
  }
  \label{fgr:model}
\end{figure*}

\mysection{2. Tile displacement model} 

\noindent
The abstract model developed in this work, which we call the Single-Assembly Tile Displacement (SATiDi) model, defines the behavior of a single tile assembly within a sea of monomer tiles.  There are a number of assumptions that must hold in order for the model to be experimentally plausible, while also allowing its definition to be fairly clean.

\begin{description}
\item[Singularity.] The concentration of multi-tile assemblies is sufficiently low (e.g. there is exactly one) that they do not interact with each other.
\item[Monomers.] Binding between two monomer tiles is sufficiently weak (at the given temperature and concentrations) that any dimers are fleeting and their presence can be neglected.
\item[Stability.] Tiles within an assembly (e.g. with four neighbors) are sufficiently strongly attached that they will not dissociate; tiles on the boundaries and corners, with only two or three neighbors, may have special binding interactions that make them as stable as the internal tiles.
\item[No growth.] With a single-side attachment being unstable for dimers, similarly new tiles may not attach by a single side to a facet of a multi-tile assembly.  When the assembly is rectangular, as will be exclusively considered here, that means the number of tiles in an assembly will never change.
\item[Full toeholds.] For consistent tile displacement kinetics, we require that the displacement process on each side has its own mediating toehold, so a tile that is bound to four neighbors will be displaced by a tile that forms a toehold on each of the four sides.  While fewer toeholds may be sufficient for displacement, it is all too plausible that their kinetics would be irregular; our simulator will issue a warning whenever such a displacement possibility is encountered.
\item[Energetics.] Tile displacement reactions must be either energetically neutral or energetically downhill, i.e. the number of toeholds formed is either the same as or more than the number of toeholds broken.
\item[Uniform design.] Each side of every tile will consist of a first toehold domain, a branch migration domain, and a second toehold domain.  We will assume that the branch migration domains are distinct on the north, east, west, and south such that they force tile to maintain a specific orientation (although non-oriented versions of the model could be formulated when non-oriented tiles are desirable).   Because branch migration domains cannot be changed by tile displacement, they will not be formally represented or accounted for in the model.
\end{description}

The model is illustrated in figure~\ref{fgr:model}A, where both a valid neutral tile displacement and a valid downhill tile displacement are shown.  Neutral displacement can be though of as generalizing the ``toehold exchange'' mechanism from strand displacement~\cite{zhang2009control}: formation of mediating toehold ensures fast kinetics, while dissociation of prior toeholds both ensures that the reaction is energetically neutral and opens up those toeholds for use in subsequent steps, as shown in figure~\ref{fgr:model}B.
Toehold exchange in tile displacement was demonstrated experimentally~\cite{petersen2018information}, although not in the exact geometric configuration required here; tuning of toehold strength (e.g. via temperature) would be required to ensure that the dissociation step (which may involve breaking four toeholds simultaneously) is sufficiently fast while still being effective for mediating the reaction.

Formally, a SATiDi system is defined by (1) a finite set of square tile types $\mathcal{S}$, each of which specifies an ordered pair of bond types (toeholds) for each of the four sides, (2) a bond strength function for each bond type $b$, $E_b > 0$, (3) a concentration for each tile type $i$, $c_i$, and (4) a standard tile displacement rate constant $k$.
The associated set of assemblies $\mathcal{A}$ consists of finite arrays of tile types (or $empty$).
Given a specific assembly, we say that a specific toehold on a specific tile is {\it closed} if the corresponding toehold on the neighboring tile has the same (i.e. matching) bond type (i.e. they form a bond), while we say that it is {\it open} otherwise.
The {\it bond energy} $E(A)$ of an assembly $A$ is the sum $\sum -E_b$ over all closed toeholds in the assembly, while the {\it free energy} $G(A)$ of the assembly is its bond energy plus the sum $\sum \ln c_i/c_0$ over all tiles in the assembly, where $c_0$ is the reference concentration (e.g. $1$~M).
Given these, we associate a formal chemical reaction network (CRN) with reactions
$$A + t_i \overset{k}{\longrightarrow} A' + t_j$$
where $A$ is an assembly with tile $t_j$ at some position $x$, $A'$ is the same assembly but with $t_i$ instead at that same position $x$, and $t_i$ is a {\it valid displacement}: on all sides where $t_j$ has a neighbor, $t_i$ forms a matching bond with (at least one) open toehold, and the total number of matching bonds increases or stays the same (i.e. the assembly's bond energy decreases or stays the same).
When all reactions are reversible, which implies that the bond energy of the assembly never changes, the CRN satisfies detailed balance with respect to the assembly bond energy, with monomer tiles having zero energy.

We consider standard stochastic kinetics according to Gillespie simulation with chemostatted constant monomer tile concentration~\cite{gillespie2007stochastic,schmiedl2007stochastic}.
For an initial state containing a single assembly, this results in a finite continuous-time Markov chain (CTMC) where the set of states are all assemblies reachable via tile displacement reactions, and the transition $A \rightarrow A'$ involving invading tile $t_i$, as above, will have rate $k \times c_i$.
If all reactions are reversible, this CTMC will satisfy detailed balance with respect to the assembly free energy, such that the equilibrium probability of assembly $A$ is
$$p(A) = \frac{1}{Z} e^{-G(A)} \hbox{\ \ \ \ \ with \ \ \ \ \ }  Z = \sum_{A'} e^{-G(A')} $$
where the partition function sum $Z$ is taken with respect to all assemblies reachable by tile displacement.

A tile displacement system simulation is considered {\it unreliable} if at any time there is an energetically neutral or downhill tile replacement that does not form at least one new toehold with each neighboring tile.  In this case, the simulation issues a {\it warning}, as illustrated in figure~\ref{fgr:model}A.

We briefly consider possible experimental implementations of single-assembly tile displacement systems.  Figure~\ref{fgr:model}C shows the motivating DNA origami tile scheme, using a geometrically-symmetrical tile design modeled after those used in several prior experimental works~\cite{tikhomirov2017programmable,tikhomirov2017fractal,petersen2018information,aghebat2020barcoded,mishra2021photocontrolled}.
More speculatively, in figure~\ref{fgr:model}D we envision an implementation that makes use of topologically two-dimensional arrays of single-stranded tiles~\cite{yin2008programming,wei2012complex}, which have been shown to tolerate a wide variety of structural variations (including single-stranded regions as we would require for toeholds) and permitting strand displacement reactions that remove tiles from the array~\cite{wei2013design,wei2014complex}.
However, single-strand tile reactions analogous to the four-toehold reversible tile displacement reactions required here have not been experimentally demonstrated.
Regardless of whether considering DNA origami tiles or single-stranded tiles -- or something else -- a major obstacle to any experimental implementation would be the creation of the initial array with a desired initial pattern.
One possible avenue -- still difficult -- would be to initially assemble a uniquely-addressed DNA origami array~\cite{tikhomirov2017fractal} or single-stranded tile array~\cite{wei2012complex}, use that array to geometrically organize the desired pattern of non-uniquely-addressed tiles needed for tile displacement behaviors, and then via photocleavable bonds or other mechanisms, remove and dispose of the uniquely-addressed array.
But for now, we will assume that arbitrary initial assemblies can be synthesized.

\begin{figure*}[tb!]
\centerline{\includegraphics[scale=0.3]{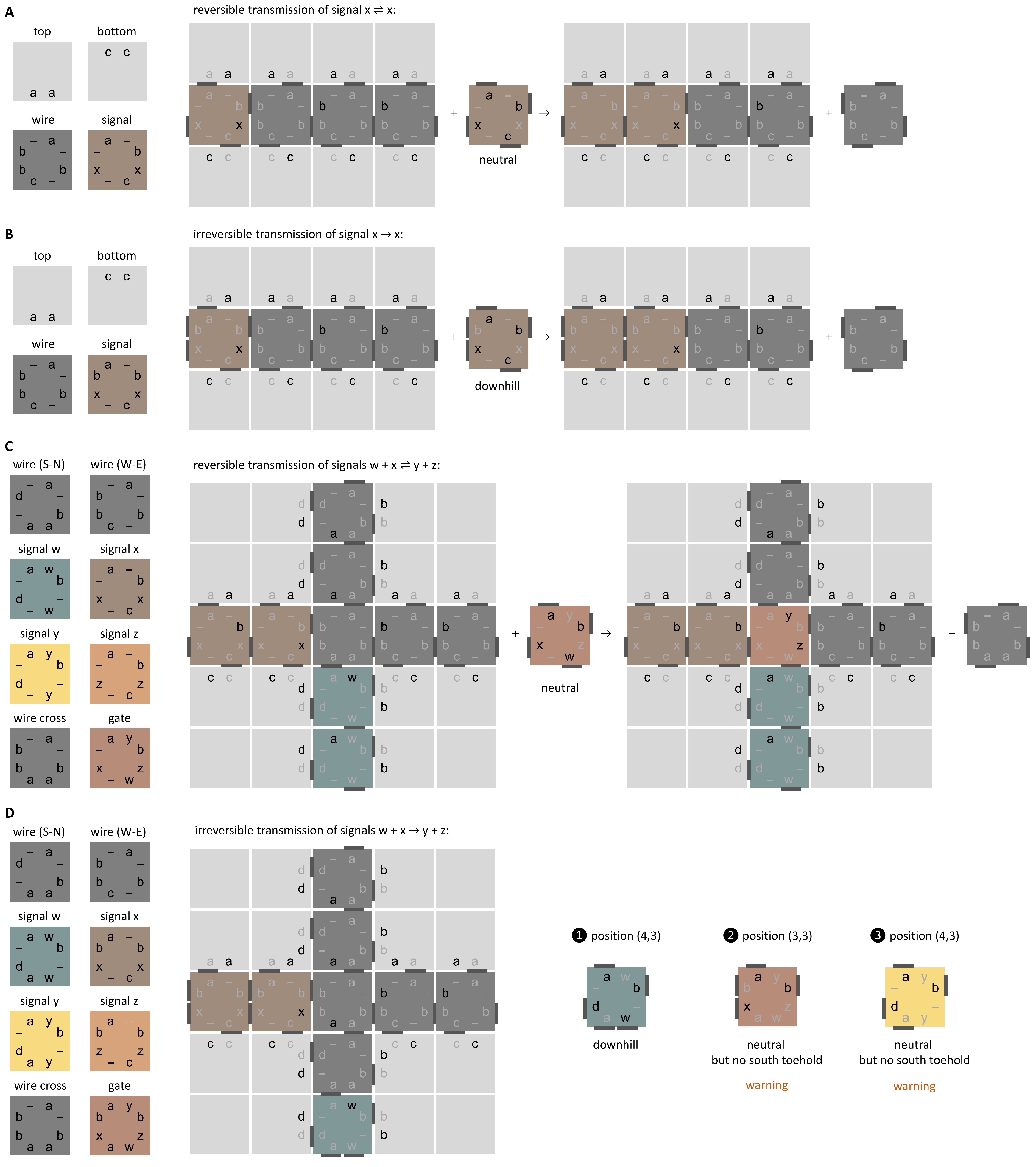}}
  \caption{\small {\bf }
  {\bf (A)} Reversible wire.  All toeholds are strength $1$ except for toehold ``$-$'', which is inert, i.e. strength $0$.  The top and bottom tiles have toeholds that are not shown, such that in the assembly the unlabeled sides are bound to each other via matching closed toeholds.  In assemblies, closed toeholds are shown with light grey labels and a solid dark grey bar indicates their bond.   For open toeholds, black or light grey is used to highlight relevant locations for the tile displacement reaction of interest, but have no formal meaning.  In the monomer tile that is a reactant of the indicated reaction, solid dark grey bars indicate where new toehold bonds will be formed. Here and in later figures, only the forward reaction is shown for any reversible reactions (i.e. the assembly and monomer tile that are the products of the indicated reaction do not have their relevant toeholds highlighted for the backward reaction).
  {\bf (B)} Irreversible wire.
  {\bf (C)} Reversible wire cross.  The initial assembly shown here illustrates the moment when both reversible signals arrive at the wire cross location.  At this time, a reversible bond-energy neutral reaction can occur that inserts the gate tile in the central location, enabling reversible signals $y$ and $z$ to propagate on the output wires.
  {\bf (D)} An unreliable irreversible wire cross that has two possible types of warnings.
  }
  \label{fgr:wire}
\end{figure*}

\mysection{3. Wires, gates, and circuits}

\noindent
To get a feel for how tile displacement systems can be programmed, we begin with the most basic task: signal transmission.
As shown in figure~\ref{fgr:wire}A, this can be accomplished using a single tile type (``wire'') that is used in the initial assembly to indicate where the wire is, plus a single tile type (``signal'') that carries the signal $x$.
Two additional tile types (``top'' and ``bottom'') are used to provide neighboring tiles for the wire, as in general the wire will be embedded within a larger assembly.
Each tile displacement reaction is neutral with respect to the bond energy, so when both the wire and signal tiles are at the same concentration, every tile displacement occurs at the same rate, and the signal transmission performs an unbiased random walk.  Thus the expected time for signal transmission along a wire of length $N$ is $O(N^2)$.

Faster signal transmission is possible if each tile displacement step is irreversible, which can be accomplished if new toehold bonds are formed such that the bond energy change is downhill.
Shown in figure~\ref{fgr:wire}B, the wire is as before, but now the signal tile has an additional toehold.  Thus, tile displacement reactions are energetically downhill, forming one net additional bond with each reaction step, and the expected time for signal transmission is now $O(N)$.

When a horizontal and a vertical wire meet, we can perform a computational step.
Figure~\ref{fgr:wire}C shows two reversible wires, one carrying signal $x$ and the other carrying signal $w$, meeting at a ``wire cross'' tile in the center.  At this location, reversible tile displacement by a ``gate tile'' can effect the $w + x \rightleftharpoons y + z$ reaction.
Because the initial wire cross tile has four closed toeholds, tile displacement by the gate tile must form all four new toehold bonds, and thus tile displacement here prior to arrival of both the $x$ and $w$ signals would be energetically unfavorable and would not occur.
This gate design is robust and flexible:  it is straightforward to design more powerful variants. For example, the horizontal wire can carry one of two signals, $0$ or $1$, the vertical wire also can carry $0$ or $1$, and there are now four gate tiles, one for each input combination, with output signals that effectively compute the logic function of interest.
Specifically, to compute NAND and output using the same signal varieties, we would use four gate tiles that replace $(w,x,y,z)$ respectively by $(0,0,1,1)$, $(0,1,1,1)$, $(1,0,1,1)$, and $(1,1,0,0)$.

Can we similarly perform logic gate operations using irreversible wires, thus making computation faster?
Unfortunately, the above schemes no longer work in this case, as illustrated in figure~\ref{fgr:wire}D.
The problem is that now, prior to arrival of the second signal, an energetically neutral tile displacement is possible at the gate position that simply ignores the missing input wire.
Indeed, if the vertical wire is meant to be capable of carrying two signals (here $w$ or $y$), then an energetically neutral tile displacement could analogously flip the signal content.
Thus, this tile set and gate design is deemed unreliable and would issue warnings in our simulator.
The lesson is that irreversible reactions will form an additional toehold when operating as intended, and this presents the possibility that a similar context that differs by just one open toehold, where the reaction is not intended to occur, will be energetically neutral and lead to an error.

\begin{figure*}[tb!]
\centerline{\includegraphics[scale=0.3]{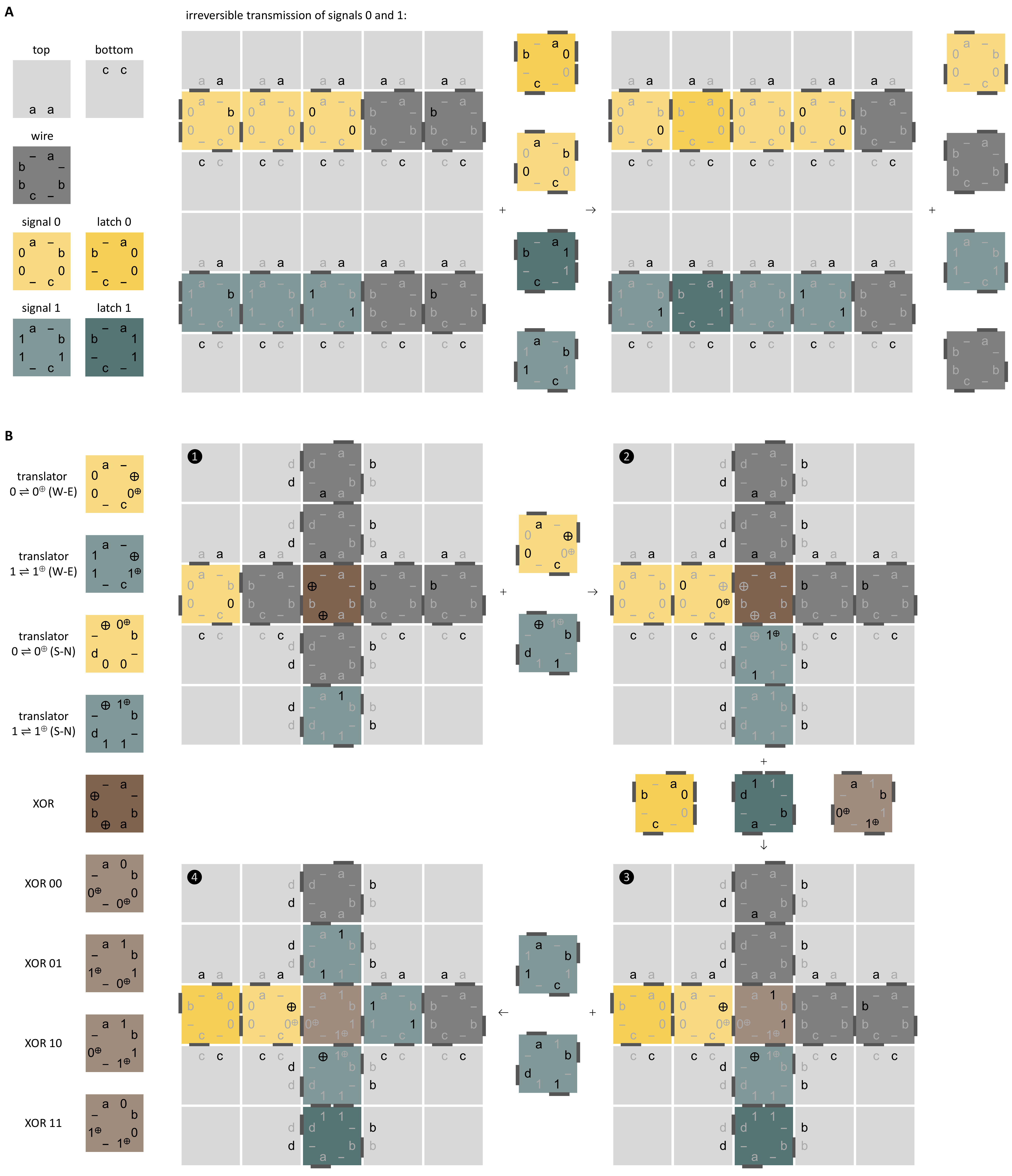}}
  \caption{\small {\bf }
  {\bf (A)} Irreversible wire with no warnings.  The next effect of two separate, independent tile displacement steps is shown for each wire.   Importantly, no unreliable tile displacement reactions are possible.
  {\bf (B)} An XOR gate.  The gate itself is entirely reversible; latch steps in the input and output wires are sufficient for ensuring net progress within a circuit.
  }
  \label{fgr:logic}
\end{figure*}

So does this mean that linear-time binary signal transmission and circuit computation is impossible with tile displacement systems?
Thankfully, no.
The trick is that while the leading wavefront of signal propagation and computation still must be reversible, in order to reliably discriminate single-toehold differences, it can be safe to irreversibly latch a decision in a context where all neighboring tile contain the same information, so differences between a 0 signal and a 1 signal by necessity involve two toeholds.
Now, if the irreversible tile displacement involves the formation of one extra toehold in the correct context, in the incorrect context it would have to ignore two toeholds and thus would be uphill.
This principle is illustrated in the design shown in figure~\ref{fgr:logic}A, where the latch tile can irreversibly insert itself into a three-tile-long segment of signal-carrying wire.
Exactly where the latch tile inserts does not matter; the signal ratchets forward either way.

We are now ready to take these designs for wires and gates, and combine them to construct feedforward logic circuits that compute in time linear with the depth of the circuit (as laid out in an array).
There is, however, one more problem to solve if we want to build circuits that utilize multiple types of logic gates (e.g. XOR, AND, OR, NAND, NOR, WIRECROSS, and others).  When an invading gate tile (e.g. ``XOR 10" shown in figure~\ref{fgr:logic}B) displaces the initial gate tile (e.g. ``XOR"), it makes bonds with toeholds on the neighboring four tiles but not with the displaced tile itself -- therefore, information about which function should be computed must be contained in the neighboring tiles, and not just the gate tiles.
We achieve this goal by using a gate-specific toehold in the initial gate tile, which directs the incorporation of a translator tile in the final position of each wire, as shown in figure~\ref{fgr:logic}B.
Now the translator tile contains information about which logic function should be computed.
Thus, an arbitrary number of gate types may coexist in the same system.

\begin{figure*}[tb!]
\centerline{\includegraphics[scale=0.3]{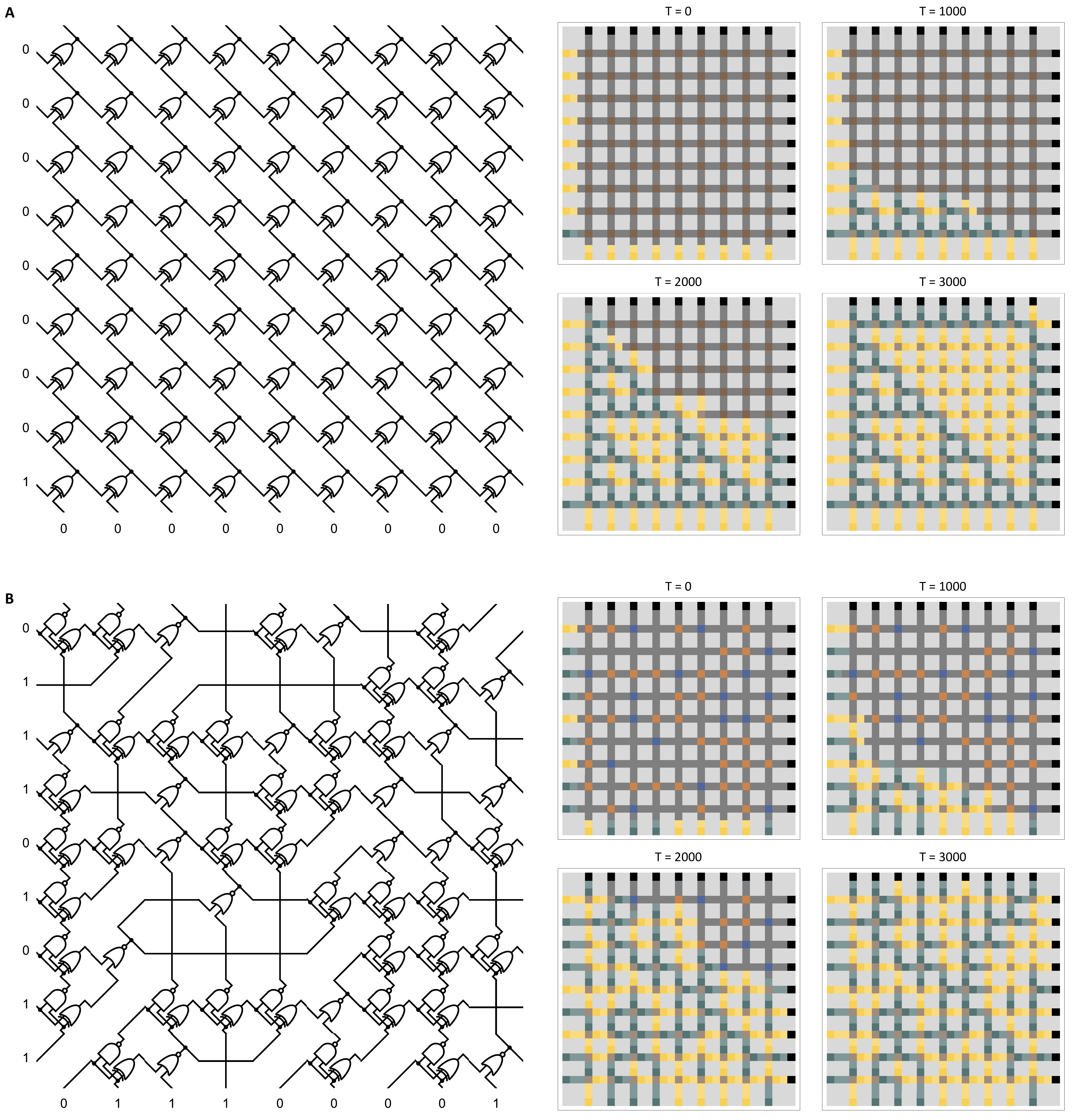}}
  \caption{\small {\bf }
  {\bf (A)} A 9~by~9 array of XOR gates.  Black tiles are ``caps'' that terminate the output wires.
  {\bf (B)} A 18-input 18-output logic circuit composed of XOR, NAND, and NOR gates. Two types of wire routing are implemented with a WIRECROSS tile that sends it south input to the north and its west input to the east, and a WIREPASS tile that sends its south input to the east and its west input to the north.  The simulation time $T$ measures the number of tile displacement reactions that have occurred, rather than the real time in the Gillespie simulation.
  }
  \label{fgr:circuit}
\end{figure*}

Simulations of two feedforward circuit computations are shown in figure~\ref{fgr:circuit}.  Using just a systolic array of XOR gates, a collection of parity outputs (involving different subsets of the inputs) are produced, incidentally creating a Sierpinski triangle pattern within the completed wires and gates.
The second circuit makes use of four gate types: a NOR gate that produces the same signal on both output wires, a NAND/XOR gate that produces NAND to the north and XOR to the east, a WIRECROSS that sends it south input to the north and its west input to the east, and a WIREPASS that sends its south input to the east and its west input to the north.
The positions and identities of the gates are laid out in the initial tile array.
Computation of an $N \times N$ circuit will take expected time $O(N)$.

The tile system with eastward and northward latching binary wires and the five types of gate functions discussed above consists of 66 tile types altogether.
For a circuit that can be laid out effectively in this format, an area of $O(N^2)$ tiles can support $N^2$ gates.
Arbitrary feedforward circuits with $N$ gates can be implemented in $O(N^2)$ area using a standard crossbar array architecture (for example see \cite{brailovskaya2019reversible}) and a new FANOUT gate that copies one input and ignores the other (using 75 tiles types if the new gate is just added, or 57 tile types if the redundant NAND/XOR and XOR gates are removed).

Can we do better than just feedforward circuits?
It is clear from inspection that our latching binary-signal wires can transmit information in either direction, depending on where the signal first arrives, and it is straightforward to implement gates that receive inputs from any two sides and produce outputs on the two other sides, so we can arrange for signals to go around in cycles.
Furthermore, the tile displacement model in principle allows displacement to occur an arbitrary number of times in a given location.
As a trivial example, the reversible wire of figure~\ref{fgr:wire}A can endlessly perform a random walk, back and forth forever.
This raises the prospect of a tile displacement system of size $N$ simulating a recurrent (iterated, feedback) circuit of size $O(N)$, which can perform computations of length $2^{O(N)}$ -- exponentially more than what a feedforward circuit of the same size can do.
This is to say, with respect to the size of the initial array and tile set, our existing construction can solve $PTIME$ problems, while a reach goal would be to solve $PSPACE$ problems like recurrent circuits can.
Unfortunately, this is not compatible with the use of latching wires to ensure linear-time signal propagation: an area-$N$ tile array initially has at most $O(N)$ open toeholds, and thus at most $O(N)$ irreversible tile displacement steps can take place before the system comes to a standstill -- or more precisely, until it must henceforth rely exclusively on reversible steps.

\mysection{4. 1D CA space-time histories}


\noindent
Boiling down what we learned about circuits to its reversible essence, we can re-implement the above computations using fewer tile types, more compact layouts with just one tile per logic gate, and power for driving the computation forward coming from concentration differences rather than from irreversible toehold formation.

\begin{figure*}[tbp!!!]
\centerline{\includegraphics[scale=0.3]{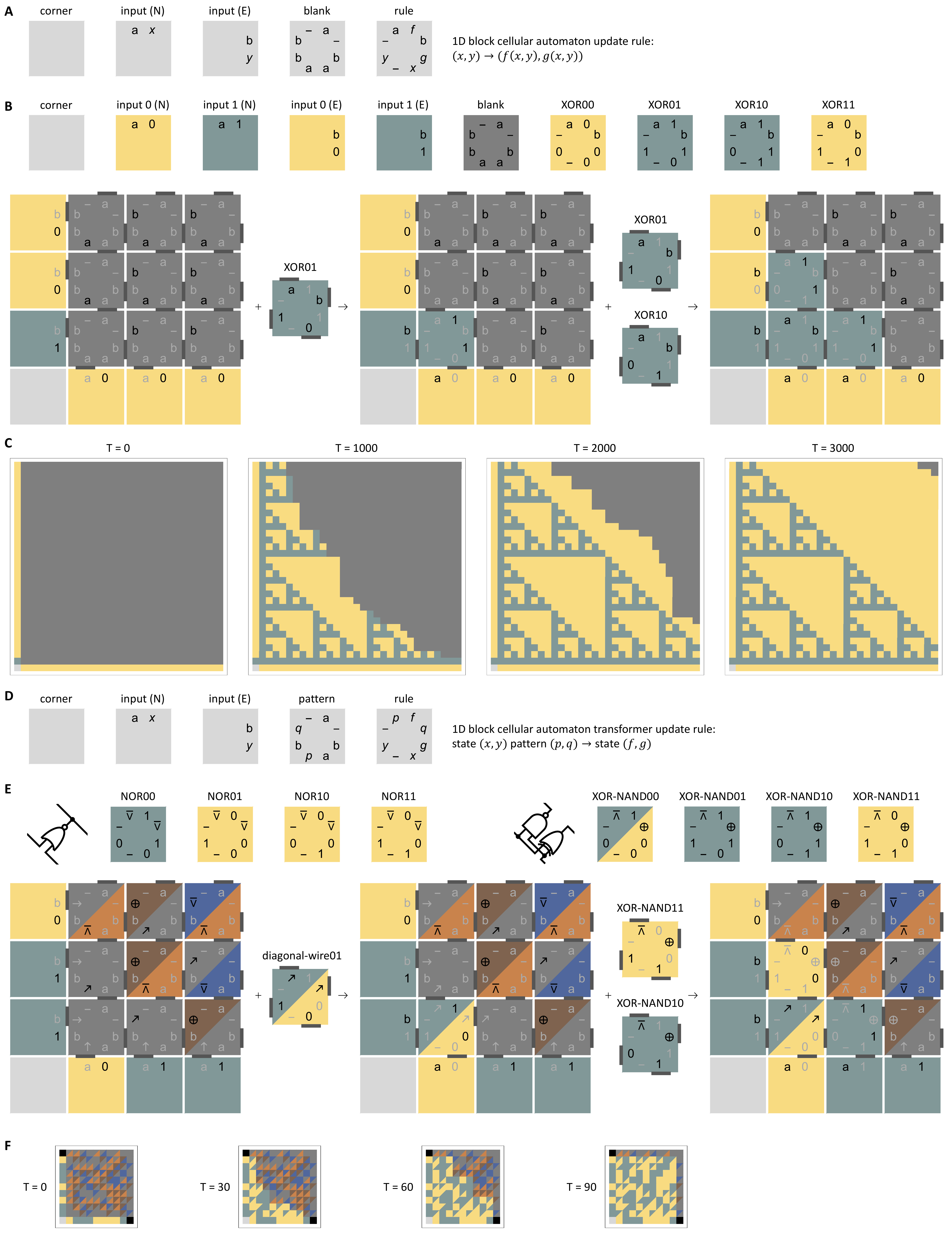}} \vspace{-.3cm}
\caption{\small {\bf }
  {\bf (A)} General case implementation of 1D block cellular automaton.  Here ${\rm a}$ and ${\rm b}$, written in roman font, denote specific toeholds. In contrast, $x$, $y$, $f$ and $g$ are variables and thus shown in italics.  There will be a separate rule tile for each possible pair $x,y \in \mathcal{A}$, with $f$ and $g$ being dependent on $x$ and $y$, and similarly for the input tiles.
  {\bf (B)} An example 1D block cellular automaton that computes the same function as the circuit shown in Fig.~\ref{fgr:circuit}A.
  {\bf (C)} Simulation snapshots.
  {\bf (D)} General case implementation of 1D block cellular automaton transformer.
  {\bf (E)} An example 1D block cellular automaton transformer that computes the same function as the circuit shown in Fig.~\ref{fgr:circuit}B.
  {\bf (F)} Simulation snapshots.
  }
  \label{fgr:1DCA}
\end{figure*}

We start by providing generalized construction for simulating the space-time history of one-dimensional block cellular automata (1D BCA) that is very similar to their simulation by algorithmic self-assembly of DNA tiles~\cite{winfree1996computational,doty2012theory}.
The instantaneous state of a 1D BCA is just a one-dimensional array of symbols from a given alphabet $\mathcal{A}$, and in each time step the entire array is synchronously updated by applying a rule $(x,y) \rightarrow (f(x,y),g(x,y))$ to a partitioning of the array into pairs, where $f$ and $g$ are functions that define the BCA and the parity of the partition alternates on each time step.
The size of the array may be infinite, finite, or expanding, with given initial state and boundary conditions (typically a finite core then periodic).
Our tile displacement system construction, shown in figure~\ref{fgr:1DCA}A, makes use of $2 + 2 N +N^2$ tile types for a 1D BCA with an alphabet of size $N$.
The initial array uses $1$ tile in the lower-left corner, $N$ tiles to define input boundary conditions to be fed in at each time step from the left, $N$ tiles to define the input boundary conditions to be fed in at each time step from the bottom, and $1$ tile type filling in the remaining ``blank'' uncomputed region of the array.
The remaining $N^2$ tile types encode every input/output case for the update rule.
For example, a binary alphabet ($N=2$) will result in 10 tile types (figure~\ref{fgr:1DCA}B).  
The $n^{th}$ synchronous update of the 1D BCA will be encoded in the $n^{th}$ diagonal of the tile array.
Similar to the gate tiles in the circuit construction, displacement must match all four open toehold positions, else it will be energetically unfavorable.
This can only happen when both the tile to the left and the tile below have already updated, thus ensuring that the computed information is based on the correct information from the preceding diagonal.

Because our model insists that any tile that can be displaced in a simulation must have a non-zero concentration as a monomer in solution, every reaction will be reversible.
However, by chemostatting the blank tile at a lower concentration than the rule tiles, each displacement reaction can be biased forward by some factor $r = c_{\rm rule}/c_{\rm blank}$.
From detailed balance of the CRN and CTMC, this ensures that the equilibrium probability of the rule-tile containing assembly is $r$ times higher than that of the blank-tile containing assembly.
Although the system will never get irreversibly locked into a final output assembly state, the complete assembly with all rule tiles in place will be $r^m$ times more likely than an assembly with $m$ blank tiles still present, which we consider ``good enough''.
Note that if a final irreversible step is desired to lock in place the completed computation, this is also possible by adapting the techniques used in the circuit construction, just at the upper right corner.

Comparing the circuit construction of figure~\ref{fgr:circuit}A to the cellular automaton space-time history construction in figure~\ref{fgr:1DCA}BC, both of which compute parallel systolic arrays of XOR gates, we see that for the same size array, the cellular automaton approach computes roughly $9$ times more gates.
It also uses just $10$ tile types, compared to $30$ for the circuit construction (if the tiles used for logic gates other than XOR are omitted).

However, our cellular automaton construction, by its very nature as a cellular automaton, receives information only in the initial 1D boundary conditions, and thus an assembly cannot specify a two-dimensional layout for the circuit that will be computed by tile displacement.
A simple modification of the ideas resolves this apparent limitation: we generalize the construction to {\it cellular automaton transformers} whose cell update now depends both on the current state $(x,y) \in \mathcal{A}_1$ and a time-and-space-dependent input pattern $(p,q) \in \mathcal{A}_0$, as shown in figure~\ref{fgr:1DCA}D.
Instead of an initial array containing uniform blank tiles, the initial array will contain a layout of ``pattern'' tiles that each encode the information $p$ that the gate below it will need to read, as well as the information $q$ that the tile to its left will need to read. If $\mathcal{A}_1$ is size $N$ and $\mathcal{A}_0$ is size $M$, then there are $2 N$ input tiles, $M^2$ pattern tiles, and $N^2 M^2$ rule tiles.
Each reversible tile displacement reaction now must match {\it four} variable pieces of information, in two pattern toeholds and two state toeholds.
As shown in figure~\ref{fgr:1DCA}EF, laying out exactly the same circuit as in figure~\ref{fgr:circuit}B now requires $9$ times less space, uses just 39 tile types ($N=2$ state bit values plus a terminator, $M=5$ logic functions, but not all combinations are needed) instead of 57, and, with concentration bias again, computes significantly faster.

Both these constructions exhibit strong similarities to computation via algorithmic growth during self-assembly of tiles -- in the first case, 2D tiles growing a 2D structure from a 1D boundary~\cite{winfree1996computational,doty2012theory}, and in the second case, 3D tiles growing an additional layer on top of a patterned 2D initial assembly~\cite{lin20103d}.
A significant difference is that rather than growing in size, the tile displacement system always remains the same size; rather than each tile attachment requiring new bond energy to counteract the lost entropy due to localization of the tile, the tile displacement system remains neutral with respect to bond energy because each incoming tile is balanced by an outgoing tile.
Thus, rather than finding suitable operating conditions by balancing temperature (controlling the bond energies) against tile concentrations (which simultaneously affect the kinetics), in tile displacement we balance concentration against concentration (which permits similar bias at different speeds and temperatures).
These benefits reflect similar observations about the increased robustness of strand displacement and toehold exchange compared to direct hybridization of complementary oligonucleotides~\cite{zhang2012optimizing,tang2020dna}. 
Seen more generally, tile displacement systems involve reconfiguration of a constant-sized assembly via local propagation of information, which is reminiscent of the distinction between crystal growth from monomers in dilute solution (the case generally assumed in algorithmic self-assembly of DNA tiles) versus crystallization from the melt (wherein the initial state is a disorganized constant-density liquid of monomers, within which crystalline order locally propagates during crystal growth).

Have we identified new concepts for tile displacement systems that allow us to perform more computation in a limited space?
Powering computation forward via concentration bias in reversible reactions has given rise to compact constructions that naturally avoid the unreliability warnings that plagued our initial wire and circuit constructions, but the computational power still remains PTIME.
One way of looking at this is that the free energy of the assembly, $G(A)$ decreases every time a higher-concentration tile replaces a lower-concentration tile, yet the minimum (most favorable) free energy occurs if all tiles in the array are highest-concentration tiles.
That is to say, the free energy is bounded below, and if each forward computational step is biased by a minimum amount, there are a bounded number of such steps that can occur before the computation is done.
The situation is not so different from the limitation we encountered when powering computation by new toehold formation in irreversible displacement steps.
Is this limitation to PTIME a feature of tile displacement systems in general, or is it particular to the lack of imagination in the constructions we have presented so far?

\mysection{5. 2D CA in-place execution}

\noindent
We can get some ideas from the notion of a cellular automaton transformer, which reads a 2D pattern as a wave of activity passes over it, leaving a new pattern in its wake.
Suppose that the new pattern can be read by a second wave, corresponding to a second cellular automaton transformer using a new set of rule tiles.
For example, the initial pattern might use toehold alphabet $\mathcal{A}_0$, the first cellular automaton transformer uses states in alphabet $\mathcal{A}_1$ and writes a new pattern using alphabet $\mathcal{A}_2$ by utilizing the two locations that, in figure~\ref{fgr:1DCA}D, have useless inert ``$-$'' toeholds.
Then, the second cellular automaton transformer can read $\mathcal{A}_2$, store its transient state in $\mathcal{A}_3$, and write a third pattern using $\mathcal{A}_4$.
To drive the computation forward, the first transformer's rule tiles should have a higher concentration than the pattern tiles, and the second transformer's rule tiles should have a higher concentration than the first transformer's rule tiles.
This idea could be extended to $K$ waves, each with its own set of rule tiles.
This would improve upon the previous constructions, in which each location in the array experiences just net one forward tile displacement step -- at that location, either one has the initial tile, or the final tile.
Whereas, in an implementation of a multiple-wave cellular automaton transformer, each location would go through a sequence of changes, one for each wave.
In a sense, we achieve $K$-fold more computation within the same assembly area.
This is somewhat analogous to freezing cellular automata, which are restricted to change a cell's state a limited number of times~\cite{chalk2018freezing}.

There are two problems here, as you have probably already noticed.
First, if the concentration ratio from wave to wave is $r$, then a $K$-wave computation requires a ratio of $r^K$ between the lowest-concentration tiles and the highest-concentration tiles.
That quickly becomes impractical, and theoretically unappealing.
Second, each wave requires a new set of tiles -- yet for PSPACE computations we would require an exponential number of tile updates and thus a comparable number of waves.
So this idea doesn't get us where we want to go.

To keep a constant number of tile types while allowing an unbounded number of tile displacement steps per site, perhaps we could have a small number $K$ of waves, but have wave $K$ output its new pattern using alphabet $\mathcal{A}_0$ so that the tiles of wave $1$ can read it -- thus allowing iterated computation, such as binary counters and perhaps universal space-bounded algorithms.
This is indeed the essence of the construction we'll arrive at, but it comes at a cost: for wave $1$ tiles to displace wave $K$ tiles, they cannot be at a lower concentration, which basically implies that all rules tiles must be at the same concentration, and we have no concentration bias pushing the computation forward.
(This conclusion is not specific to periodic waves of cellular automata transformers; it follows in general that if we want to implement a computation that may update a given site an unknown and unbounded number of times, then every tile type may at some point be an incoming tile and at other times be the outgoing tile, so the concentrations of all rule tiles must be equal.)
If we have already accepted that our designs should exclusively use bond-energy neutral tile displacement, then in fact the bond energy and free energy of our assembly will remain constant over time -- we are truly dealing with reversible computation.
Thankfully, reversible computation is by no means impossible~\cite{bennett1982thermodynamics,morita2017theory}.

\begin{figure*}
\centerline{\includegraphics[scale=0.3]{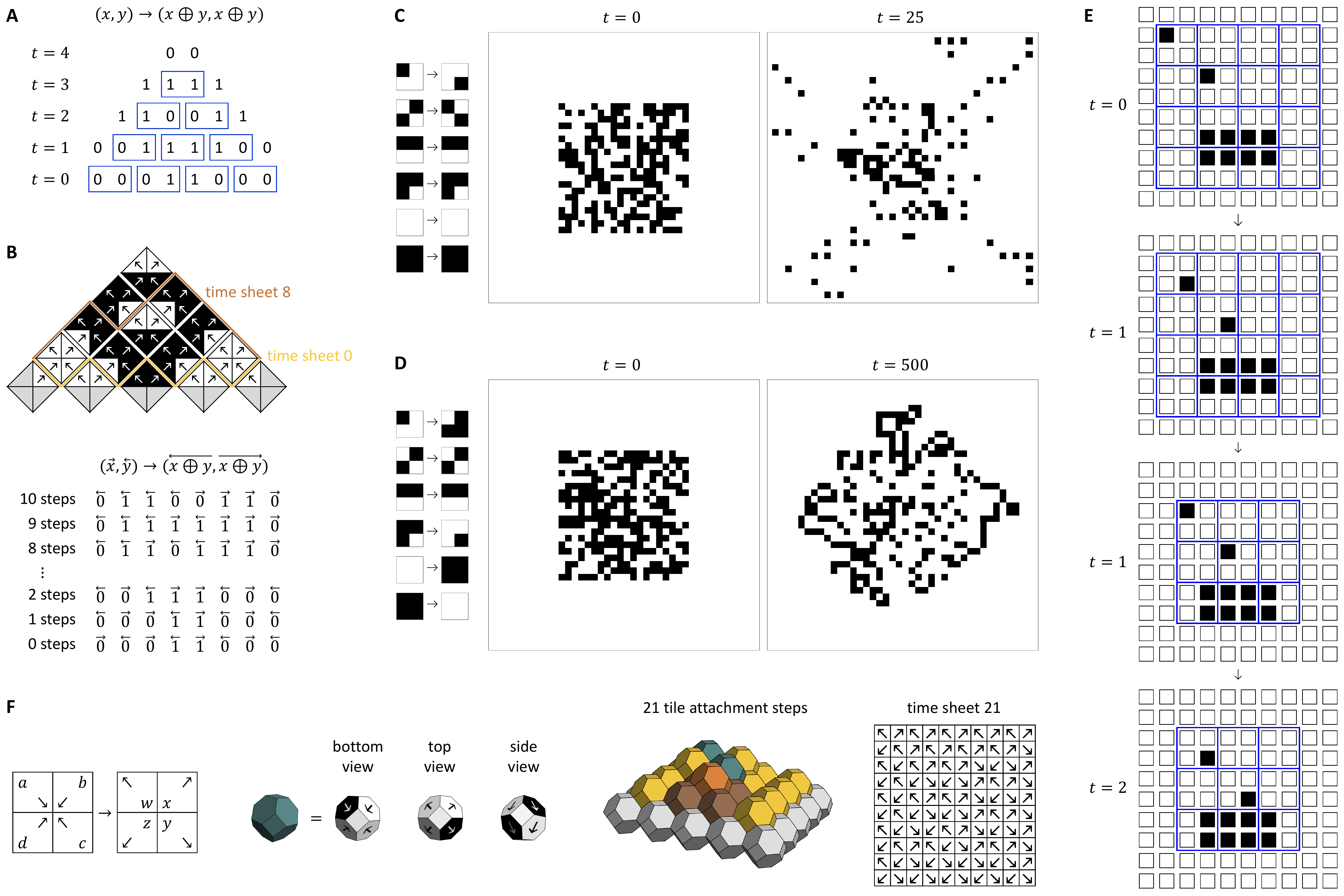}}
  \caption{\small {\bf }
  {\bf (A)} Execution of a synchronous 1D block cellular automaton.
  {\bf (B)} Asynchronous 2D tile self-assembly that simulates the computation in (A).
  {\bf (C)} and {\bf (D)} Simulations of two example 2D block cellular automata: Billiard Ball Model~(C) and Critters~(D).
  {\bf (E)} Execution of a synchronous 2D block cellular automaton.
  {\bf (F)} Asynchronous 3D tile self-assembly that simulates the computation in (E).
  }
  \label{fgr:CA}
\end{figure*}

Our approach will be to exhibit a surprisingly natural correspondence between certain tile displacement systems and the well-studied class of two-dimensional block cellular automata (2D BCA) that arose in the study of reversible computation by discrete models of ballistic physical dynamics~\cite{margolus1984physics,toffoli1987cellular,margolus1999crystalline}.
The 2D BCA model is a natural generalization of the 1D BCA discussed above: rather than partitioning a 1D array into pairs of cells that get synchronously rewritten with alternating partition parity on alternate time steps, we now partition a 2D array into $2 \times 2$ blocks of cells that get synchronously rewritten with alternating partition parity on alternating time steps (compare figure~\ref{fgr:CA}A with figure~\ref{fgr:CA}E).
The formalism allows the rewrite rules to be arbitrary functions
$$f \left(\begin{bmatrix} a & b \\ d & c \end{bmatrix} \right) = \begin{bmatrix} w & x \\ z & y \end{bmatrix}$$
but if the rewrite function is a bijection, then the 2D BCA is logically reversible in the sense that iterating with $f^{-1}$ instead of with $f$ will bring the simulation backwards in time.
The most famous 2D BCA rule, the Billiard Ball Model (BBM), is logically reversible, rotationally and mirror symmetric, conserves the total number of 1s, can directly simulate reversible circuits, and with an infinite periodic initial state can simulate universal Turing machines~\cite{margolus1984physics}.
Example simulations of two binary-state 2D BCA, the BBM and ``Critters'', are shown in figure~\ref{fgr:CA}CD.
With larger alphabets, 2D BCA can simulate arbitrary classical cellular automata and Turing machines, either of the irreversible or reversible variety.
(Generalizations to using blocks larger than $2 \times 2$ is also natural, but will not be considered here.)

There are three obstacles to implementing arbitrary 2D BCA as tile displacement systems, and we will solve them all.
The first is that tile displacement reactions are asynchronous (occurring at random locations and in random orders) while 2D BCA require synchronous updates of the entire array (and fail utterly if the same update function is applied asynchronously with no other modifications).
The second is that the mechanics of tile displacement must be designed to avoid irreversible steps that close too many toeholds at once.
And the third obstacle is that with exclusively reversible reactions and no concentration bias, there must be some other way to drive the system forward if we don't want to wait forever.

For the first challenge, we adapt prior methods for imbuing asynchronous cellular automata with locally synchronizing mechanisms~\cite{lee2002reversible,lee2016characterization,clamons2020programming}.
The specific approach used here generalizes the approach used for simulation of 1D cellular block automata space-time histories in the previous section.
Figure~\ref{fgr:CA}A gives an example of a 1D BCA, with boxes highlighting the partitioning into pairs with alternate parity on each synchronous time step.
Figure~\ref{fgr:CA}B shows the same computation interpreted as 2D tile self-assembly where, starting from the 5 tiles at the bottom that encode the 8 input bits as well as their partitioning, rule tiles attach whenever they can match two sides of existing tiles in the assembly, thus asynchronously growing the space-time history.
We have augmented the tiles with arrows that point to where incoming tiles could attach; thus, in the initial assembly of 5 tiles, the sites where tile can attach are exactly those locations where arrows are pointing inward toward the incoming tile.
A cut through the assembly's space-time diagram corresponds a particular moment during the asynchronous self-assembly process -- we show a cut after 0 tile additions (yellow) and another after 8 tile additions (orange).
We call these ``time sheets'' because at different horizontal ($x$) positions, they are at different heights ($t$), and thus reading out the binary (black/white) states along a time sheet path correspond to states at different time steps of the underlying synchronous cellular automaton.
Nonetheless, the time-sheet state information, augmented with the relevant arrows, is all that is needed to correctly complete the computation using an asynchronous update rule that executes only when arrows point toward each other, otherwise leaving the cells untouched.
This process exactly mimics the self-assembly of the deterministic space-time history, despite its non-deterministic order of execution.

There is an exactly analogous arrow-augmented asynchronous update rule for 2D BCA.
Rather than square tiles, we now have truncated octahedra as ``tiles'', but the self-assembling structure is again a space-time history of the correct synchronous cellular automaton computation.
Tiles may attach when they match four hexagonal faces of existing tiles in the assembly.
(The small square faces are inert.)
Again, if we imagine arrows orthogonal to the hexagonal faces of tiles, pointing out of the tile, then valid sites for attachment of a new tile correspond exactly to situations where all four arrows on the matching faces are pointing toward each other.
The growth front for a give stage of assembly again corresponds to a (now two dimensional) time sheet, and we can write out the states of each exposed hexagonal face in a two dimensional array along with the orientation of its corresponding arrow.
It is now a simple observation that the asynchronous addition of a tile corresponds exactly to an asynchronous update of a $2 \times 2$ block with four inward-pointing arrows, resulting in updates of the four cells and reversing all four arrows.
Another way of thinking of it is that after a block asynchronously updates, it will not be able to update again until all four overlapping $2 \times 2$ blocks have first updated and flipped the arrows back.
Thus, the arrow-augmented asynchronous updating corresponds exactly to synchronous parallel updating with alternating-parity partitioning into blocks.

\begin{figure*}
\centerline{\includegraphics[scale=0.3]{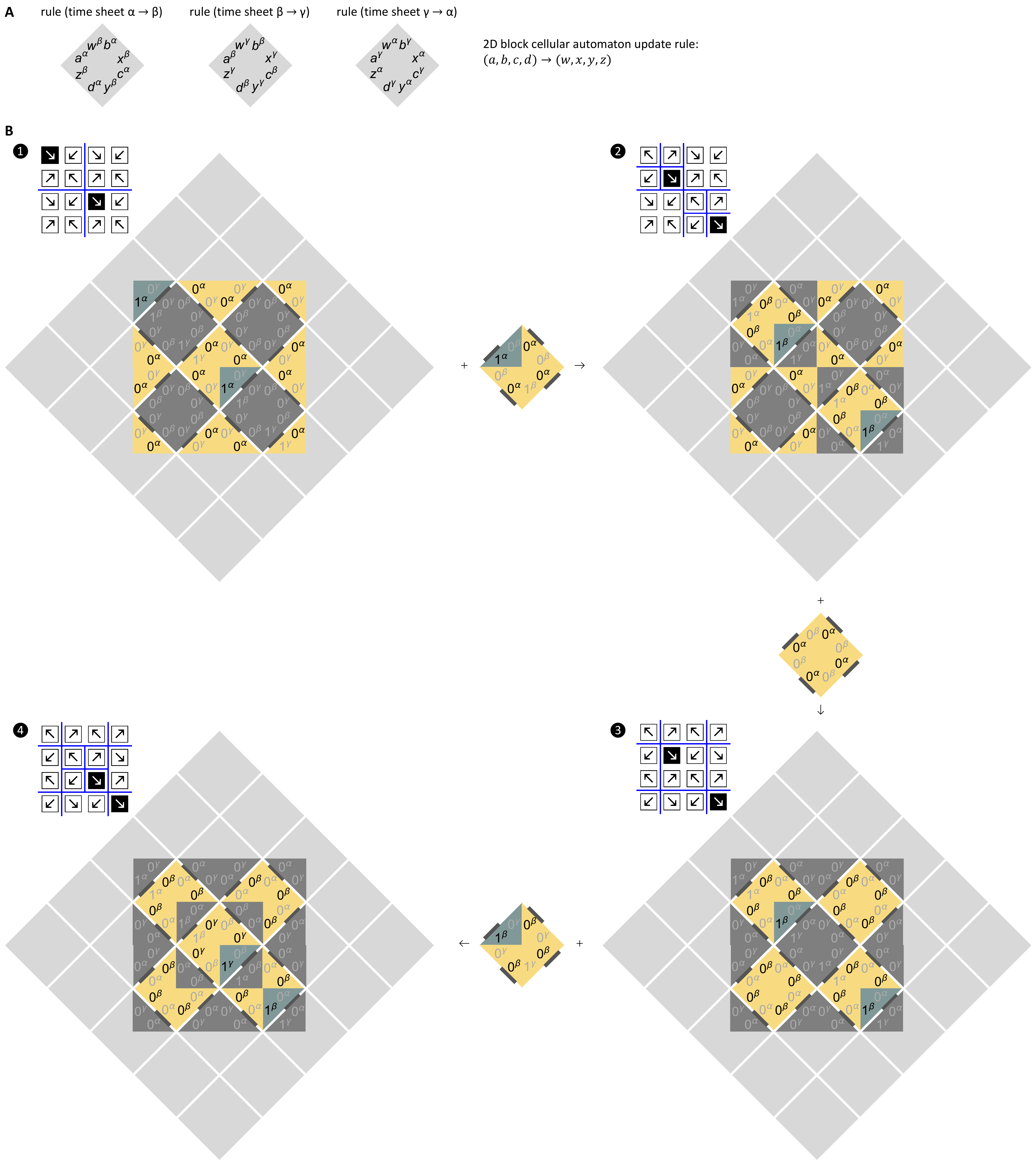}}
  \caption{\small {\bf }
  {\bf (A)} General case implementation of 2D block cellular automaton.
  {\bf (B)} Example updates in the Billiard Ball Model.
  }
  \label{fgr:2DCA}
\end{figure*}

The second challenge is to implement this type of asynchronous block cellular automaton updating rule using tile displacement.
Our construction, shown in figure~\ref{fgr:2DCA}, introduces additional complexities due to the fact that all tile displacement reactions are physically reversible, even if the 2D BCA logic update rules are irreversible, combined with the need to ensure that when one toehold is closed, the neighboring toehold must be opened -- thus we must be able to guarantee a mismatch.
Tripling the cell state alphabet by adding $\alpha$, $\beta$, $\gamma$ markers solves both problems.
For each 2D BCA update case, we make three tiles, one inputting $\alpha$-symbols and outputting $\beta$-symbols, another inputting $\beta$-symbols and outputting $\gamma$-symbols, and the third inputting $\gamma$-symbols and outputting $\alpha$-symbols.
When an $\alpha \rightarrow \beta$ tile inserts into the array, that simulates a forward-time asynchronous update.
The swapping of which toehold is open and which is closed reflects the flipping orientation of arrows in the asynchronous cellular automaton; we can read the arrows from a tile array by looking at the open toeholds and drawing the arrow from $\alpha$ to $\beta$, from $\beta$ to $\gamma$, or from $\gamma$ to $\alpha$.
Boundary conditions for finite arrays must also be handled, using the same principles.

Each side of a tile encodes the state of a specific cell in the 2D BCA (at a particular time mod 3, as per $\alpha$-$\beta$-$\gamma$ of the open toehold), and thus the grid of simulated BCA cells is oriented at a $45^\circ$ angle relative to the array of tiles.
State being encoded on the sides of tiles also facilitates that each tile displacement step corresponds to an update of a whole $2 \times 2$ block, and the fact that two tiles share the same side location reflects that each cell in a 2D BCA can be updated either by an odd-parity block or an even-parity block.

The final challenge concerns how to drive the computation forward.
Let us first consider reversible 2D BCA rules.
In this case, after any forward tile displacement step, there is exactly one monomer tile type that can reverse the reaction: the tile that was just displaced.
What this means is that the full state space of the tile displacement system's CTMC is essentially linear; though fat and fuzzy, it has the same thickness both arbitrarily far into the future and arbitrarily far into the past.  The thickness has to do with all the possible contours of the time sheet for a given average time.
Thus we can say that the state space of the tile displacement system consists exclusively of correct reachable states of the computation; for a reversible 2D BCA simulating a compact recurrent circuit for solving a PSPACE problem, the tile displacement systems's state space will also be exponentially long and will reach the same correct conclusion.
Stochastic Gillespie simulation of the tile displacement CRN will result in an unbiased random walk back and forth along this fuzzy-linear state space.
(Every assembly in this reachable state space has the same energy.)
However, unlike a standard reversible Turing machine with Brownian dynamics~\cite{bennett1982thermodynamics}, whose state spaces is strictly a linear graph so the expected random walk hitting time for reaching the end of an $T$ step computation is $O(T^2)$, the time sheet diffuses much more slowly.
As a rough estimate for an $N \times N$ tile array that requires $N^2$ forward updates to move the time sheet 1 net synchronous update step into the future under ideal circumstances, the same $N^2$ updates if half forward and half backward will be expected to net move the time sheet $N$ steps either forward or backward, which corresponds to just $1/N$ equivalent synchronous update steps.
This being just a polynomial inefficiency, perhaps we should not be too concerned.

More interesting is what happens if the 2D BCA rules are irreversible.
This means there are multiple cases for the $2 \times 2$ block input that map to the same output.
Therefore the state space for the tile displacement system will be exponentially branched in the backwards-in-time direction (as pictured by Bennett in figure 10 of his review paper~\cite{bennett1982thermodynamics}, but thicker and fuzzier).
Consequently Brownian dynamics will tend to be entropically biased toward where there are more states, and the system will run backwards.
Can this entropic driving force be used to encourage a system to perform a desired computation by designing a system whose {\it reverse} dynamics are what we want?
Attempting to do so would be risky, and probably futile, because the 2D block update rule being irreversible means that there are some states that have no local predecessor, and backward progress will get stuck as such local configurations are encountered.

A better way to exploit an entropic driving force is to have nondeterministic, stochastic forward update rules added to an otherwise-reversible system.
For example, consider a tile displacement simulation of the BBM model, with boundary tiles designed to implement a reflecting boundary (as they must for the system to remain reversible).
If we design a special boundary tile that can either reflect a ball or (in the forward direction) produce a new ball out of nothing, then we obtain a new system that is still entirely reversible in the sense that there exists (at least one) possible applicable block update in all circumstances, so the system cannot get stuck either in the forward direction or the reverse direction.
With uniform tile concentrations, all assembly states will still be isoenergetic.
But started with an empty $N \times N$ array, forward updates of the special tile will about half the time produce a new ball, which will entropically drive the system to a density such that forward production of balls is balanced by the reverse reaction, the absorption of balls into the special tile.
At this point, which will be $O(N^2)$ synchronous time steps in the future, the time sheet will stop advancing on average.
Another way of looking at it is that with all reactions being neutral, equilibrium will reflect equipartition among all reachable states, and the combinatorially greatest number of states will have a number of ball near the optimal density -- so, that's what we are likely to observe.
And the only way to get there is to run the time sheet forward enough to emit that number of balls.

A gas-filled BBM simulation is not of great use by itself, but we can make use of it by also placing a circuit in the array, and drawing a 2-cell-thick wall around it.
In the BBM model, balls bounce off walls, and walls are stable.
Thus, despite random stochastic gas entering the areas of the array outside the box, the circuit will remain perfectly isolated from the gas.
But due to the time-sheet coupling enforced by the asynchronous arrow rules, the time sheet that is being driven forward by the expansion of the gas will simultaneously drive the circuit forward.

Unfortunately, for an array of area $N^2$, we will only drive the computation forward by $O(N^2)$ steps -- this is no better than the PTIME computational power of the original circuit construction.
Essentially, in a small confined space, our circuit ``heats up'' and stops working.
To run it for a long time, we need a larger space into which we can release the simulated heat.
For example, if we are willing to entertain a half-infinite-plane array for tile displacement, we can draw a BBM wall down the middle, release gas on one side, and let the other side simulate an interesting recurrent circuit.
Now, although the array is infinite (or very large) in direct proportion to how much computation we want to do, we can say that we have confined the interesting part of the computation -- the circuit itself -- to a very small area relative to the potentially exponentially long computation.
This isn't PSPACE computation in terms of the size of the array, but rather in terms of the size of the part of the array that we care about.
Similar constructions can be used to drive forward computation not just for other reversible 2D BCA rules, but even for irreversible rules:
the cellular automaton alphabet can be expanded to encode an inert ``wall'', time sheets within the walled region and outside of it remain coupled, and sufficient entropy must be generated by stochastic rules outside the wall, to be dissipated into a sufficiently larger area.


\mysection{6. Discussion}

\noindent
Tile displacement within arrays of square DNA origami tiles was discovered accidentally~\cite{petersen2018information}.
While some aspects of the formal model, such as the four-sided generalization of toehold exchange, were invented for mathematical elegance rather than detailed realism, they are not too far flung from what has been experimentally demonstrated and characterized.
So it is quite delightful that within the design space for tile displacement systems, we find natural implementations for feedforward circuits and one-dimensional cellular automata that compute in linear time, powered by irreversible toehold formation or by concentration gradients.
Even more delightful is that attempts to squeeze out more computational power per area seemed almost inevitably to lead us consider physical constraints such as energy, reversibility, and asynchrony -- which in turn lead to classical two-dimensional cellular automata models that arose in early studies of the physics of computation~\cite{bennett1982thermodynamics,margolus1984physics}.
Our strongest result (despite weak time efficiency) is that a tile displacement array of size $N$ can reversibly simulate a recurrent reversible circuit (via the Billiard Ball Model cellular automaton) for an arbitrary number of steps.
In other words, the reachability question for tile displacement is PSPACE complete -- a result strongly reminiscent of Thachuk \& Condon's beautiful PSPACE-hardness result for CRNs and DSDs~\cite{thachuk2012space}.

We believe that there remains a lot undiscovered within the tile displacement design space.
For example, while our constructions showed that the asynchronous tile displacement model can simulate synchronous cellular automata, the needed flipping-arrow mechanism for local synchronization seems almost built-in to the tile displacement model in the form of open and closed toeholds for toehold exchange, and it's not obvious how to directly simulate asynchronous cellular automaton models such as reversible surface CRNs~\cite{qian2014parallel,brailovskaya2019reversible,clamons2020programming}.
We might also ask whether using information within the branch migration domains rather than just in toeholds -- or whether having even more toeholds and branch migration domains on a tile's sides -- could have advantages either theoretically or experimentally.
Further, the most interesting systems demonstrated experimentally in the initial work on tile displacement~\cite{petersen2018information} involved systems of interacting multi-tile arrays, rather than a single array and a monomer.
How do our single-assembly results fit into that larger picture?
Finally, might tile displacement systems be combined with other molecular mechanisms to solve our problems driving the computation forward -- for example, an oscillator~\cite{srinivas2017enzyme} that periodically activates and deactivates the $\alpha$, $\beta$, and $\gamma$ monomer tiles in sequence.

When first discovered, the tile displacement mechanism seemed most closely related to strand displacement mechanisms, only two dimensional.
However, as we investigated the capabilities of single-assembly tile displacement, many parallels to passive tile assembly~\cite{doty2012theory} became prominent.
Tile displacement systems appear to combine the principles of DNA strand displacement and self-assembly in different ways than hairpin-based programmable self-assembly~\cite{yin2008hairpins}, signal-passing tile self-assembly~\cite{padilla2014asynchronous,padilla2015signal}, CRN-controlled tile assembly~\cite{zhang2013integrating,schiefer2015universal}, and other models we are aware of.
Comparing the benefits, drawbacks, and relationships between these models may help uncover a more unified way of thinking about programmable molecular systems.

And even if tile displacement systems, as explored theoretically here, never become useful experimentally, we hope that it was interesting and perhaps inspiring to look long and deep at a simple mechanism until intricate patterns emerge.

\mysection{Acknowledgements}

\noindent
The authors thank William Poole 
and Ho-Lin Chen 
for useful discussions.
This work was partially supported by NSF awards 2008589 and 1813550.

\vspace{-0.2cm}
\renewcommand{\refname}{}
{\footnotesize
\bibliographystyle{unsrt}
\bibliography{tile_displacement_theory_ref}

\end{document}